\documentclass{aastex62}

\renewcommand{\baselinestretch}{1.5} 

\graphicspath{{./}{figures/}}

\submitjournal{ApJ}
\received{October 23, 2019}
\revised{December 20, 2019}
\accepted{December 20, 2019}

\begin{document}

\title{Constraints on nonlinear tides due to $p$-$g$ mode coupling from the neutron-star merger GW170817}

\correspondingauthor{Steven Reyes}
\email{sdreyes@syr.edu}

\author[0000-0002-4599-6054]{Steven Reyes}
\affil{Syracuse University,
Syracuse, NY 13244, USA}

\author[0000-0002-9180-5765]{Duncan A. Brown}
\affiliation{Syracuse University,
Syracuse, NY 13244, USA}
\nocollaboration



\begin{abstract}
It has been suggested by \cite{Weinberg:2013pbi} that an instability due to the nonlinear coupling of a neutron star's tide to its $p$- and $g$-modes could affect the gravitational-wave phase evolution of a neutron-star binary. \cite{Weinberg:2015pxa} suggests that this instability can turn on as the gravitational waves pass through the sensitive band of ground-based detectors, although the size of the effect is not known. The discovery of the binary neutron star merger GW170817 provides an opportunity to look for evidence of nonlinear tides from $p$-$g$ mode coupling. We compute Bayesian evidences that compare waveform models that include the $p$-$g$ mode coupling to models that do not. We find that the consistency between GW170817 signal and the $p$-$g$ mode model 
reported by \cite{abbott2019constraining} is due to a degeneracy between the 
phenomenological waveform used to model the effect of 
nonlinear tides and the standard post-Newtonian waveform. We investigate the consistency of the GW170817 signal with regions of the parameter space where the effect of nonlinear tides is not degenerate with the standard model. Regions of the nonlinear tide parameter space that have a fitting factor of less than 99\% (98.5\%) are  disfavored by a Bayes factor of 15 (25). We conclude that regions of the parameter space where nonlinear tides
produce a measurable effect are strongly disfavored and improved theoretical
modeling will be needed if future observations are to constrain nonlinear
tides from $p$-$g$ mode coupling in neutron stars.
\end{abstract}

\keywords{binaries: close - stars: neutron - stars: oscillations}


\graphicspath{{./}{figures/}}

\section{Introduction} \label{sec:intro}
The discovery of the binary neutron star merger GW170817 \citep{TheLIGOScientific:2017qsa} has given us a new way to explore the physics of neutron stars. Recent studies have measured the star's tidal deformability and placed constraints on the equation of state of the neutron stars~\citep{TheLIGOScientific:2017qsa,Tews:2018iwm,Most:2018eaw,Raithel:2018ncd,de2018tidal,Abbott:2018exr,Abbott:2018wiz,Radice:2018ozg,LIGOScientific:2019eut,Capano:2019eae}.
\cite{Weinberg:2013pbi} have suggested that the star's tidal deformation can induce nonresonant and nonlinear daughter wave excitations in $p$- and $g$-modes of the neutron stars via a quasi-static instability. This instability would remove energy from a binary system and possibly affect the phase evolution of the gravitational waves radiated during the inspiral. Although \cite{Venumadhav:2013nla} concluded that there is no quasi-static instability and hence no effect on the inspiral, \cite{Weinberg:2015pxa} claims that the instability can rapidly drive modes to significant energies well before the binary merges. However, the details of the instability saturation are unknown and so the size of the effect of the $p$-$g$ mode coupling on the gravitational waveform is not known~\citep{Weinberg:2015pxa}. The discovery of the binary neutron star merger GW170817 by Advanced LIGO and Virgo provides an opportunity to determine if there is evidence for nonlinear tides from $p$-$g$ mode coupling during the binary inspiral.

Since the physics of the $p$-$g$ mode instability is uncertain, \cite{Essick:2016tkn} developed a parameterized model of the energy loss due to nonlinear tides. This model is parameterized by the amplitude and frequency dependence of the energy loss, and the gravitational-wave frequency at which the instability saturates and the energy loss turns on. For plausible assumptions about the saturation, \cite{Essick:2016tkn} concluded that $> 70\%$ of binary merger signals could be missed if only point-particle waveforms are used, and that neglecting nonlinear tidal dynamics may significantly bias the measured parameters of the binary. Bayesian inference can be used to place constraints on nonlinear tides during the inspiral of GW170817. An analysis by \cite{abbott2019constraining} computed Bayes factors that investigate whether the GW170817 signal is more likely to have been generated by a model which includes nonlinear tides or one which does not. \cite{abbott2019constraining} find a Bayes factor of order unity, and conclude that the GW1701817 signal is consistent with both a model that neglects nonlinear tides and with a model that includes energy loss from a broad range of $p$-$g$ mode parameters. However, the prior space used in this analysis includes a large region of  parameter space where the amplitude of the effect produces a gravitational-wave phase shift that is extremely small. In this case, a waveform that includes $p$-$g$ mode parameters will have a likelihood that is identical to the likelihood of the waveform without the $p$-$g$ mode instability. The $p$-$g$ mode model extends the standard waveform model by adding additional parameters that describe the nonlinear tidal effects. However, when including new parameters in a hypothesis if the likelihood does not vary across large portions of the prior volume for these new parameters relative to the likelihood of the original model, then the Bayes factor will not penalize this additional prior volume, nor will it penalize any extraneous parameters in the model (see e.g. \cite{kass1995bayes,hobson2010bayesian}). We examine the prior space of the $p$-$g$ mode model used by \cite{abbott2019constraining} and find that although the $p$-$g$ model model contains regions that are not consistent with the standard model, there are large regions of the prior space where the likelihood is high because the $p$-$g$ mode model is degenerate with the standard model. These regions of the prior space dominate the evidence and hence the Bayes factor neither favors nor disfavors the inclusion of $p$-$g$ mode parameters.

We investigate a variety of different prior distributions on the $p$-$g$ mode parameters beginning with a prior distribution that is similar to that tested in~\cite{abbott2019constraining} and includes large regions of the parameter space that produce a negligible gravitational-wave phase shift. When comparing the evidence for this model with the standard waveform model used by \cite{de2018tidal} we find a Bayes factor of order unity, as expected. We then investigate a prior distribution in which the $p$-$g$ mode instability parameters are constrained to induce a phase shift to the waveform that is greater than $0.1$ radians. This phase shift is calculated from the time the waveform enters the sensitive band of the detector to the time when the waveform reaches the innermost stable circular orbit. We choose this threshold to exclude trivial regions of the parameter space that produce a non-measurable effect. However, we again find a Bayes factor of order unity when compared to the model hypothesis that does not model the $p$-$g$ mode instability. Investigation of these results showed that this is due to parameter degeneracies between the $p$-$g$ mode model and the intrinsic parameters of the standard waveform model.

Finally, we reduce the prior space to contain only the regions where the $p$-$g$ mode waveform is not degenerate with the standard model by computing the fitting factor~\citep{Apostolatos:1995pj} of $p$-$g$ signals against a set of standard waveforms. We do this to restrict the region of parameter space to that where the $p$-$g$ effect is \emph{measurably} distinct from a model that neglects nonlinear tides. We calculate the Bayes factor as a function of the fitting factor. We find that as the $p$-$g$ mode parameter space is restricted to exclude regions that have a high fitting factor with standard waveforms, the Bayes factor decreases significantly. Regions of the nonlinear tide parameter space that have a fitting factor of less than 99\% (98.5\%) are strongly disfavored by a Bayes factor of 15 (25).  While certain prior distributions of $p$-$g$ mode parameters are consistent with the data, we find that these distributions are ones that contain large regions of non-measurable parameter space either because the effect produced is too small to measure, or the effect is degenerate with other parameters of the standard model. We conclude that the consistency of the GW170817 signal with the model of \cite{Essick:2016tkn} is due to degeneracies and that regions where nonlinear tides produce a measurable effect are strongly disfavored.

\section{Waveform model}
\label{sec:waveform}

As two neutron stars orbit each other, they lose orbital energy $E_\mathrm{orbital}$ due to gravitational radiation $\dot{E}_{GW}$. The gravitational waveform during the inspiral is well modeled by post-Newtonian theory~(see e.g. \cite{Blanchet:2013haa}).  The effect of the $p$-$g$ mode instability is to dissipate orbital energy by removing energy from the tidal bulge of the stars~\citep{Weinberg:2013pbi,Weinberg:2015pxa,Essick:2016tkn}. Once unstable, the coupled $p$- and $g$-modes are continuously driven by the tides, giving rise to an extra energy dissipation $\dot{E}_{NL}$ for each star in the standard energy-balance equation~\citep{Peters:1963ux} 
\begin{equation}
\dot{E}_\mathrm{orbital} = -\dot{E}_\mathrm{GW} - \dot{E}^1_\mathrm{NL} - \dot{E}^2_\mathrm{NL}.
\label{eqn:energy_bal}
\end{equation}
Since the details of how the nonlinear tides extract energy from the orbit is not known, \cite{Essick:2016tkn} constructed a simple model of the energy loss and calculated plausible values for the model's parameters. In this model, the rate of orbital energy lost during the inspiral is modified by 
\begin{equation}\label{eqn:energy_nl}
\dot{E}_\mathrm{NL} \propto A f^{n+2} \Theta (f - f_0),
\end{equation}
where $A$ is a dimensionless constant that determines the overall amplitude of the energy loss, $n$ determines the frequency dependence of the energy loss, and $f_0$ is the frequency at which the $p$-$g$ mode instability saturation occurs and the effect turns on. By solving Eq.~(\ref{eqn:energy_bal}), \cite{Essick:2016tkn} computed the leading order effect of the nonlinear tides on the gravitational-wave phase as a function of $A$, $n$, and $f_0$. In this analysis, they allowed each star to have independent values of $A$, $f_0$, and $n$, but found that the energy loss due to nonlinear tides depends relatively weakly on the binary's mass ratio. Hence, they consider a model that performs a Taylor expansion in the binary's component mass~\citep{DelPozzo:2013ala} and include only the leading order terms in the binary's phase evolution. Given this, we parameterize our nonlinear tide waveform with a single set of parameters $A$, $n$, and $f_0$, by setting $\dot{E}^1_\mathrm{NL} = \dot{E}^2_\mathrm{NL}$. We  keep only the leading order nonlinear tide terms when we obtain the quantities $t(f)$ and $\phi(f)$ used to compute the stationary phase approximation~\citep{Sathyaprakash:1991mt,Droz:1999qx,Lindblom:2008cm}. This approach is reasonable for GW170817, since both neutron stars have similar masses and radii~\citep{de2018tidal}.

The dependence of $A$, $n$, and $f_0$ on the star's physical parameters is not known~\citep{Weinberg:2015pxa}. \cite{Essick:2016tkn} estimate that plausible parameter ranges are $A \lesssim 10^{-6}$, $0 \lesssim n \lesssim 2$, and $30 \lesssim f_0 \lesssim 80$ Hz. \cite{Zhou:2018tvc} found that the frequency at which the instability begins to grow is equation-of-state dependent and can occur at gravitational-wave frequencies as high as $700$~Hz. \cite{Andersson:2017iav} suggest that the instability may only act during the late stages of inspiral, (above $300$~Hz), otherwise the large energy dissipation will cause the temperature of the neutron stars to be very large. 

In this paper, we compare two models for the gravitational waves radiated by GW170817. The first is the standard restricted stationary-phase approximation to the Fourier transform of the gravitational waveform $\tilde{h}(f)$, known as the TaylorF2 waveform~\citep{Sathyaprakash:1991mt}. We begin with the same waveform model used by \cite{de2018tidal}, which is accurate to 3.5 PN order in the orbital phase, 2.0 PN order in spin-spin, self-spin and quadrupole-monopole interactions, 3.5 PN order in spin-orbit coupling, and includes the leading and next-to-leading order corrections from the star's tidal deformability~\citep{Kidder:1992fr,Blanchet:1995ez,Blanchet:2004ek,Buonanno:2009zt,Arun:2008kb,Marsat:2013caa,Bohe:2013cla,Bohe:2015ana,Mikoczi:2005dn, Flanagan:2007ix,Vines:2011ud}. We then construct a second model that adds the leading order effect of nonlinear tides computed using the model of \cite{Essick:2016tkn}. We compute the Fourier phase for the TaylorF2 model $\Psi(f)_\mathrm{TaylorF2}$ and add a term that accounts for the additional energy lost due to nonlinear tides $\Psi_\mathrm{NL}(f)$, given by
\begin{equation}
\Psi_\mathrm{NL}(f) = - \frac{25}{768} A \left(\frac{G \mathcal{M} \pi f_\mathrm{ref}}{c^3} \right)^{-\frac{10}{3}} \times \left \{
                     \begin{array}{ll}
                       \left( \frac{f_0}{f_\mathrm{ref}} \right)^{n-3} \left[ \left( \frac{f}{f_0}\frac{1}{n-4} \right) - \frac{1}{n-3} \right] &\quad f < f_0, \\
                       \left( \frac{f}{f_\mathrm{ref}} \right)^{n-3} \left(\frac{1}{n-4} - \frac{1}{n-3} \right)  &\quad f \ge f_0.
                     \end{array}
                     \right.
\label{eqn:fourier_phase_eq1}
\end{equation}
Here, $f_\mathrm{ref}$ is a reference frequency which we set to $100$~Hz following \cite{Essick:2016tkn}, $G$ is Newton's gravitational constant, $c$ is the speed of light, and $\mathcal{M} = (m_1 m_2)^{3/5}/(m_1+m_2)^{1/5}$ is the chirp mass of the binary.\footnote{Appendix~A of \cite{Essick:2016tkn} gives the change to the gravitational-wave phase $\phi(f)$ as a function of frequency and not the change to the Fourier phase $\Psi(f)$ (see e.g. \cite{Lindblom:2008cm} for a discussion of how these differ). The former quantity is useful to compute the change in the number of gravitational-wave cycles, but the latter is required to compute the modification to the TaylorF2 waveform. The study by~\cite{abbott2019constraining} corrects this mistake.} This waveform model can have a degeneracy in the gravitational-wave phasing with chirp mass when $n = 4/3$. For this value of $n$, the Fourier phase in Eq.~(\ref{eqn:fourier_phase_eq1}) for nonlinear tides is $\Psi(f) \propto f^{-5/3}$, which is the same power law dependence as the chirp mass phasing. A degeneracy occurs when $f_0$ is comparable or lower than the frequency at which chirp mass can be accurately measured. In this case, the $p$-$g$ mode instability is degenerate with changing the chirp mass. In principle, there will be other degeneracies with other intrinsic parameters of the gravitational-wave signal for other values of $n$.

We generate the standard TaylorF2 waveform using the LIGO Algorithm Library~\citep{lal} and multiply this frequency-domain waveform by the term due to the nonlinear tides,
\begin{equation}
\tilde{h}_\mathrm{TaylorF2+NL}(f) = \tilde{h}_\mathrm{TaylorF2}(f) \times \exp[-i \Psi_\mathrm{NL}(f) ].
\end{equation}
The Fourier phase for the nonlinear tides is implemented as a patch to the version of the PyCBC software~\citep{alex_nitz_2018_1208115} used by~\cite{de2018tidal}. Both the standard and nonlinear tide waveform models are terminated when the gravitational-wave frequency reaches that of a test particle at the innermost stable circular orbit of a Schwarszchild black hole of mass $M = m_1 + m_2$. For the neutron star masses considered here, this frequency is between 1.4 kHz and 1.6 kHz.

\section{Model Priors} \label{sec:priors}
Bayes theorem offers a methodology for evaluating the plausibility of models relative to a given data set, and then updating these prior model beliefs with better hypotheses. Bayes theorem states that
\begin{equation}
p\left(\vec{\theta}\,| H, \mathbf{d}\right) = \frac{ p\left(\mathbf{d}|H, \vec{\theta}\,\right)\, p\left(\vec{\theta}\,|H\right)}{p\left(\mathbf{d}|H\right)},
\label{eq:bayestheorem}
\end{equation}
where $p \left(\mathbf{d}|H \right)$ is the evidence of the model $H$, $p\left(\vec{\theta}\,|H\right)$ is the prior distribution of the parameters given the signal model, $p\left(\mathbf{d}|H, \vec{\theta}\right)$ is the likelihood of the data for a particular set of parameters $\vec{\theta}$, and $p\left( \vec{\theta}\,|H, \mathbf{d}\right)$ is the posterior distribution of the parameters given the signal model. The likelihood used in this analysis assumes a Gaussian model of detector noise and depends upon the noise-weighted inner product between the gravitational waveform and the data from the gravitational-wave detectors~\citep{Finn:2000hj,Rover:2006bb}. The choice of prior distributions on the parameters of the signal model represent the hypothesis that we want to test. The posterior distributions reflect how to update ones beliefs with respect to the likelihood and the data. Thus, by examining many different parameter hypotheses we can investigate the extent to which GW170817 is accurately modeled by $p$-$g$ mode instability waveform models.

In our analysis, we fix the sky location and distance to GW170817~\citep{Soares-Santos:2017lru,Cantiello:2018ffy} and assume that both neutron stars have the same equation of state by imposing the common radius constraint~\citep{de2018tidal}. In the case of the standard TaylorF2 waveform $H_\mathrm{TaylorF2}$, our analysis is identical to that described in~\cite{de2018tidal}. This analysis considered three prior distributions on the binary's component mass. Here, we only consider the uniform prior on each star's mass, with $m_{1,2} \sim U[1,2]\, M_\odot$, and the Gaussian prior on the component masses $m_{1,2} \sim N(\mu = 1.33, \sigma = 0.09)\, M_\odot$~\citep{Ozel:2016oaf}. For both mass priors, we restrict the chirp mass to the range $ 1.1876 M_\odot < \mathcal{M} < 1.2076 M_\odot$. Since our analysis is identical to that of~\citep{de2018tidal}, we refer to that paper for the details of the data analysis configuration.

Given the uncertainty on the range of the nonlinear tide parameters, we follow~\cite{abbott2019constraining} and let $n \in U[-1, 2.999]$, draw $A$ from a distribution uniform in $\log_{10}$ between $10^{-10}$ and $10^{-5.5}$, and $f_0 \in U[10, 100]$~Hz. We use this along with a uniform prior distribution on the mass from~\cite{de2018tidal}.

\begin{figure}[th]
\centering
\includegraphics[width=0.9\columnwidth]{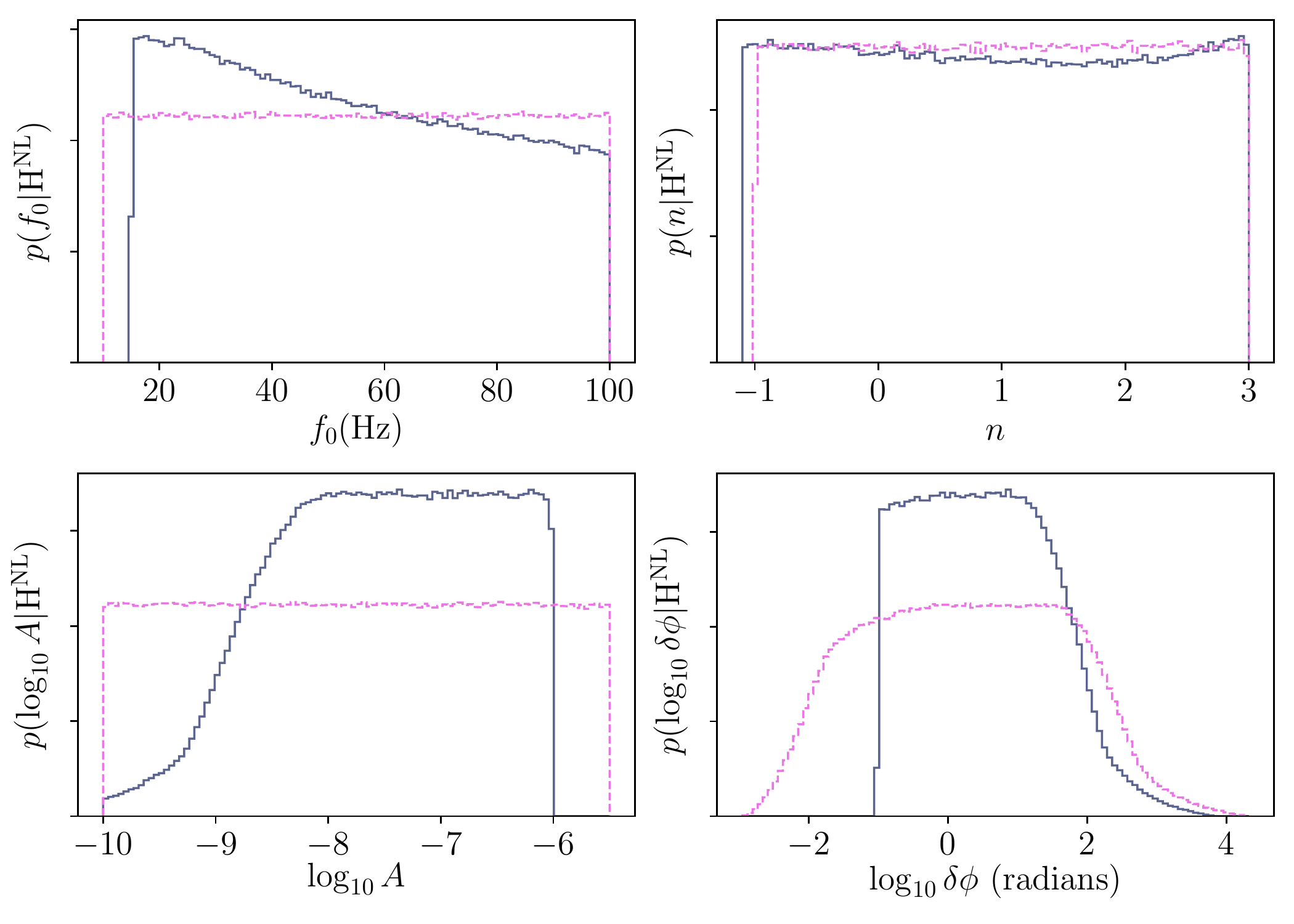}
\caption{Prior probability distributions on the parameters $(f_0, n, A)$ for the waveform model $\mathrm{H}^{\mathrm{NL}} = \mathrm{H}_\mathrm{TaylorF2+NL}$ and the resulting prior on the gravitational-wave phase shift $\delta\phi$ shift due to nonlinear tides. The dark blue, solid lines shows the priors when $f_0$ is drawn from a uniform distribution between $15$ and $100$~Hz with a $\delta\phi \ge 0.1$~rad constraint restricting some of the prior space. The pink, dotted lines represent prior distributions on the nonlinear tidal parameters similar to~\cite{abbott2019constraining}.}
\label{fig:priors}
\end{figure}

We also consider two alternative choices of drawing $f_0$: we draw $f_0$ from a uniform distribution between $15$ and $100$~Hz, as used by~\cite{Essick:2016tkn}, and from a uniform distribution between $15$ and $800$~Hz to allow for the larger values of $f_0$ suggested by~\cite{Zhou:2018tvc} and~\cite{Andersson:2017iav}. For these choices we consider $A$ uniform in $\log_{10}$ between $10^{-10}$ to $10^{-6}$. The distribution on $n$ is permitted to be $n \in U[-1.1, 2.999]$. For these alternative prior distributions we also consider applying a further constraint on the parameters. Since some combinations of $A$, $n$, and $f_0$ can produce extremely small gravitational-wave phase shifts~\citep{Essick:2016tkn}, we place a cut on the gravitational-wave phase shift due to nonlinear tides
\begin{equation}
\delta \phi(f_\mathrm{ISCO}) =
                      \frac{-25}{768}  \frac{A}{n-3} \left( \frac{G \mathcal{M} \pi f_{\mathrm{ref}}}{c^3} \right)^{-10/3} \left[ \left( \frac{f_0}{f_{\mathrm{ref}}} \right)^{n-3} - \left(\frac{f_\mathrm{ISCO}}{f_{\mathrm{ref}}} \right)^{n-3} \right],
\end{equation}
where $f_\mathrm{ISCO}$ is the termination frequency of the waveform (which is always larger than $f_0$ in our analysis). This gravitational-wave phase shift from the $p$-$g$ mode instability is strictly negative, but we take the convention of using the absolute value of the phase shift for convenience. We restrict the prior space to values of $\delta \phi > 0.1$~rad. Phase shifts of $\delta \phi \approx 0.1$~rad have an overlap between the two waveform models greater than 99.98\%. This cut means that the resulting priors on $A$, $n$, and $f_0$ are not uniform, but are biased in favor of combinations of parameters that may produce a measurable effect on the phasing of the waveform due to nonlinear tides. While $\delta \phi$ is a simple proxy for how similar or dissimilar two waveforms are, formally this is given by the match between two waveforms. A $\delta \phi$ of $1$ radian may have a low overlap with a waveform if the radian is accumulated over a large bandwidth but a high overlap if the radian is accumulated near the very end of the signal. Fig.~\ref{fig:priors} shows a depiction of the prior distributions used when using a permissive prior on $\delta \phi$, similar to~\cite{abbott2019constraining}, and when using a constraint on the $p$-$g$ mode parameters such that $\delta \phi > 0.1$~rad.

A stricter approach to constructing a prior distribution that considers $p$-$g$ mode effects that are distinguishable from standard waveforms is to examine the fitting factor between a distribution of $p$-$g$ mode waveforms and a set of comparable TaylorF2 waveforms. To do so, we examine the fitting factor of our Bayesian inference analysis with respect to a template bank of non-spinning, mass-only TaylorF2 waveforms. We construct a template bank of $\sim 20,000$ non-spinning, mass-only waveforms of comparable masses to the prior distribution on the mass parameters. The template bank is constructed with component masses, $m_{(1,2)} \in (1.0, 2.0) M_{\odot}$, chirp masses, $\mathcal{M}_c \in (1.1826, 1.2126) M_{\odot}$, and a minimal match placement of 99.9\%. We then place a threshold on the evidence calculation from the Bayesian analysis based on the maximum overlap with this template bank of standard waveforms. This permits an analysis of the Bayes factor for nonlinear tides where the prior distribution on $p$-$g$ mode parameters is determined by the fitting factor with a set of standard signals.

\section{Methods} \label{sec:methods}
We use the gravitational-wave strain data from the Advanced LIGO and Virgo detectors for the GW170817 event, made available through the GW Open Science Center~\citep{Vallisneri:2014vxa,gw170817-losc}. We then repeat the analysis of~\cite{de2018tidal} using the waveform model $H_\mathrm{TaylorF2+NL}$ to compute the evidence $p(\mathbf{d}\, | \, \mathrm{H}_\mathrm{TaylorF2+NL})$.

We use Bayesian model selection to determine which of the two waveform models described in Sec.~\ref{sec:waveform} is better supported by the observation of GW170817. Bayes theorem in Eq.~(\ref{eq:bayestheorem}) permits us a method for model  through the ratio of the evidence from each model. This ratio of the model evidences is called the Bayes factor, which we denote as $\mathcal{B}$. A Bayes factor greater than unity indicates support for the model in the numerator, while a Bayes factor less than unity indicates support for the model in the denominator. The Bayes factor can be written as,
\begin{equation}
\mathcal{B} = \frac{p\left(\mathbf{d}\,|\,\mathrm{H}_\mathrm{TaylorF2+NL}\right)}{p\left(\mathbf{d}\,|\,\mathrm{H}_\mathrm{TaylorF2}\right)}.
\label{eq:bayesfactor}
\end{equation}
The numerator of Eq.~(\ref{eq:bayesfactor}) is the evidence for nonlinear tides $p\left(\mathbf{d}\,|\,\mathrm{H}_\mathrm{TaylorF2+NL}\right)$. For the denominator of Eq.~(\ref{eq:bayesfactor}), we use the evidence $p(\mathbf{d}\,|\,\mathrm{H}_\mathrm{TaylorF2})$ provided as supplemental materials by~\citep{de2018tidal}.

Posterior distributions for parameters of interest can be also computed by marginalizing the posterior probability distribution over other parameters.  Marginalization to obtain the posterior probabilities and the evidence is performed using Markov Chain Monte Carlo (MCMC) techniques. To compute posterior probability distributions and Bayesian evidences, we use the \emph{PyCBC Inference} software~\citep{alex_nitz_2018_1208115,biwer2019pycbc} using the parallel-tempered \emph{emcee} sampler~\citep{emcee,vousden:2016}.
This sampler allows the use of multiple temperatures to sample the parameter space~\citep{emcee, doi:10.1143/PTPS.157.317, B509983H}.

From these multiple temperatures we use the thermodynamic integration method~\citep{lartillot2006computing,friel2008marginal} to estimate the logarithm of the Bayesian evidence, ln $z$, given as:
\begin{equation}
\textrm{ln} \, z = \int_0^1 \langle \textrm{ln} \, \mathcal{L} \rangle_{\beta} \, d\beta
\label{eq:thermoint}
\end{equation}
The estimate of the Bayesian evidence is determined by the integral over inverse temperatures, $\beta$, of the average log likelihood, $\langle \textrm{ln}\, \mathcal{L} \rangle_{\beta}$, for each inverse temperature $\beta$. An approximation to this integral can be made through use of trapezoid rule integration method. Following~\cite{de2018tidal} we use $51$ temperatures where we use a combination of geometric and logarithmic temperature placements to improve the accuracy of the integral~\citep{liu2016evaluating}.

We verify the results of the thermodynamic integration evidence calculation by comparing it with the steppingstone algorithm~\citep{xie2010improving}, which utilizes the same likelihoods from multi-tempering sampling as the thermodynamic integration method. Both trapezoidal rule thermodynamic integration and steppingstone methods can have some bias in the estimate of the logarithm of the Bayesian evidence due to a finite number of temperatures being used. This bias is mitigated by an increased number of temperatures~\citep{xie2010improving, Russel:2018pqv}. Additionally, this bias can be mitigated in thermodynamic integration by improving the order of the quadrature integration~\citep{friel2014improving}. We also use a higher order trapezoidal rule from~\cite{friel2014improving} and verify that the results are consistent.

We also estimate the error for each method of evidence calculation. The thermodynamic integration method and steppingstone algorithm both contain Monte Carlo error~\citep{annis2019thermodynamic}. For the thermodynamic integration method the Monte Carlo error on the thermodynamic integral can be estimated following the methodology of~\cite{annis2019thermodynamic}. We use this same uncertainty estimate for the higher order trapezoidal rule as well. In~\cite{xie2010improving} there is a Monte Carlo variance estimate for the logarithm of the evidence from the steppingstone method that we also use here.

The last source of error in the evidence calculation that we consider is whether the MCMC has converged to stable likelihood values across all of the temperatures. This requires examining the stability of the evidence calculations as the MCMC progresses. Independent samples are drawn according to the $\mathrm{n_{acl}}$ method as described by~\cite{biwer2019pycbc} at various points in the run. This method takes a specific endpoint iteration, takes half the endpoint iteration as the starting point iteration, and calculates the autocorrelation length of the samples between the starting point and the endpoint iteration. Independent samples are drawn in intervals of the maximum autocorrelation length for the samples within this segment. We divide the full run into $12$ segments and calculate the evidence from each one of these segments to examine how the evidence progresses along the MCMC iterations. Gradually the evidence begins to settle towards a constant value as the MCMC progresses. We take the difference between the last two evidence estimates as the convergence error.

We estimate the total error on our evidence calculations, $\sigma_{\mathrm{ln} \, z}$, by adding the errors in quadrature according to,
\begin{equation}
\sigma_{\mathrm{ln} \, z} = \sqrt{ \sigma_{\mathrm{MC}}^2 + \sigma_{\mathrm{convergence}}^2 } \, .
\label{eq:errorprop}
\end{equation}
Here, the error $\sigma_{\mathrm{MC}}$ is the Monte Carlo error and $\sigma_{\mathrm{convergence}}$ is the convergence error. Finally, to estimate the Bayes factors we model the log evidence as a normal distribution, with mean given from the log evidence calculation, and standard deviation given by the error propagation formula in Eq.~(\ref{eq:errorprop}). The logarithm of the Bayes factor can then be calculated from the difference in the logarithm of the evidences. The standard Bayes factor is then the exponential of the logarithm of the Bayes factor.

\section{Results}\label{sec:results}
Compared to the standard waveform model, we find that the $p$-$g$ mode model with priors where $\delta \phi$ is unconstrained gives a Bayes factor of order unity. When we use $p$-$g$ mode priors where $\delta \phi$ $>$ $0.1$~radians we also find a Bayes factor of order unity. Following the Bayes factor interpretation of \cite{kass1995bayes, jeffreys1998theory}, these Bayes factors cannot be considered to be statistically significant. A Bayes factor of unity indicates that whatever prior beliefs we had about the plausibility of the $p$-$g$ mode instability prior to GW170817 is unchanged by the observation of GW170817. For the narrow range of $15 \le f_0 \le 100$~Hz where $\delta \phi > 0.1$~rad, we find that the Bayes factors are $\mathcal{B} \sim 0.7$. This is also true of the prior range $10 \le f_0 \le 100$~Hz with unconstrained $\delta \phi$. The broader range  $15 \le f_0 \le 800$~Hz, where $\delta \phi > 0.1$~rad, we find that $\mathcal{B} \sim 0.7$ as well. Our estimated statistical error on Bayes factors due to Monte Carlo error and convergence error is $\sim \pm 0.1$ at the $90$\% confidence level.

When we consider the way that the nonlinear tides enter the Fourier phase in Eq.~(\ref{eqn:fourier_phase_eq1}), we see that if $n = 4/3$ then the nonlinear tides enter the Fourier phase of the waveform with the same power law dependence on frequency $f$ as the chirp mass, that is $\Psi(f) \propto f^{-5/3}$. We also note that for the effect of nonlinear tides to be degenerate with chirp mass, they must turn on at a frequency $f_0$ that is close to the low-frequency limit of the detector's sensitive band. If the effect turns on at higher frequencies, then the phasing will change in the detector's sensitive band and it is more difficult to compensate for the nonlinear tide effect with a change in chirp mass. 

\begin{figure}[th]
\includegraphics[width=\columnwidth]{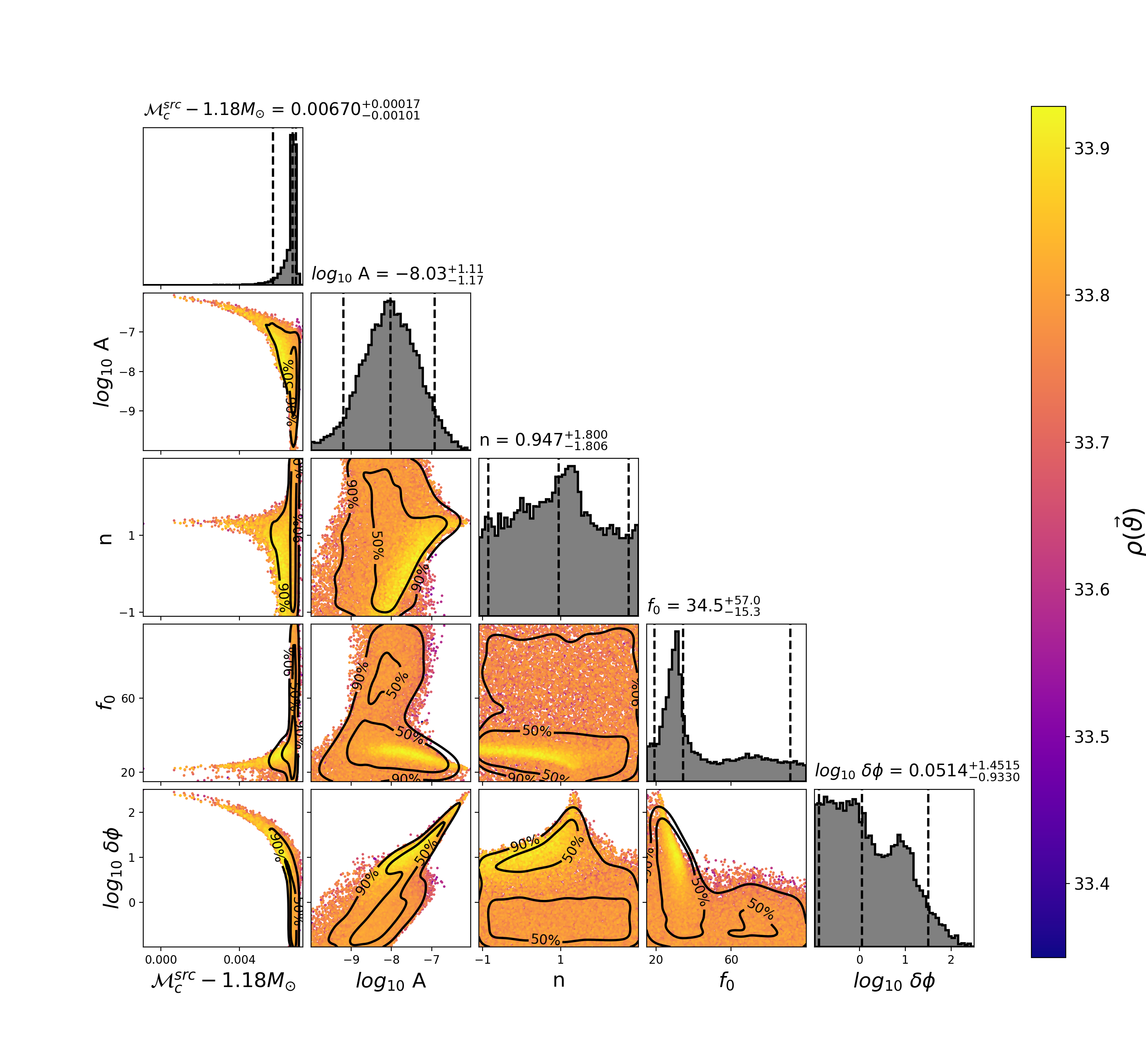}
\caption{The marginalized posterior distributions for the uniform mass prior and a $f_0$ restricted to the range $15$ and $100$~Hz. The vertical lines on the marginalized histograms display the $5$th, $50$th, and $95$th percentiles of the posteriors. The three-detector network signal-to-noise ratio for each sample is given on the color-bar. The posterior scatter plots show 50\% and 90\% credible interval contours. The posteriors on $n$ is peaked $n \lesssim 4/3$ and for values of $f_0$ close to the lower end of the detector's low frequency sensitivity. In this region of the parameter space, the effect of nonlinear tides is degenerate with chirp mass, causing a skew in the chirp mass posterior. It can be seen from the $\delta\phi$--$\mathcal{M}$ plot (lower left) that large phase shifts due to nonlinear tides are due to points in the parameter space where a value of chirp mass can be found that compensates for the phase shift of the nonlinear tides. It is notable that the peaks in the $f_0$ posterior, at $f_0 \approx 30$~Hz and $f_0 \approx 70$~Hz seem to be reversed from those in Fig 2. of \citep{abbott2019constraining}. Note that the marginalized posterior for $A$ is diminished for $A < 10^{-8}$ due to the $\delta \phi$ prior constraint.
}
\label{fig:uniform_f0_small}
\end{figure}

The marginalized posterior distributions on parameters shown in Fig.~\ref{fig:uniform_f0_small} show a strong degeneracy between the source-frame chirp mass $\mathcal{M}^\textrm{src}$ and nonlinear tides that creates a tail in the chirp mass posterior skewed towards lower values of chirp mass than the value measured using the standard waveform model, $\mathcal{M}^\textrm{src} = 1.1867\pm0.0001\, M_\odot$~\citep{de2018tidal}. We see a peak in the posteriors of $n$ and $f_0$ at $n \lesssim 4/3$ and $f_0 \lesssim 35$~Hz. This parameter degeneracy is also correlated with large $A$, where $10^{-8} \lesssim A < 10^{-6}$. The samples with large posterior values of $\delta\phi$ seen in Fig.~\ref{fig:uniform_f0_small} are strongly correlated with source-frame chirp masses $\mathcal{M}^\textrm{src} \lesssim 1.1866.$ We have examined the change to the posterior distribution when changing the low-frequency cutoff of the likelihood integration from $20$~Hz to $25$~Hz, and to $30$~Hz. In these analyses, the peak in the posterior of $f_0$ tracks the low-frequency cutoff of the likelihood integration, confirming that this effect is due to the chirp-mass degeneracy with the low-frequency cutoff. The chirp mass degeneracy is also present in the analysis with the broader range of $f_0$, however it is not as pronounced in the posterior samples due to the larger prior space being explored. For the prior distributions discussed above, the observation of GW170817 does not provide strong statistical evidence either for or against the presence of nonlinear tides. 

\begin{figure}[th]
\includegraphics[width=\columnwidth]{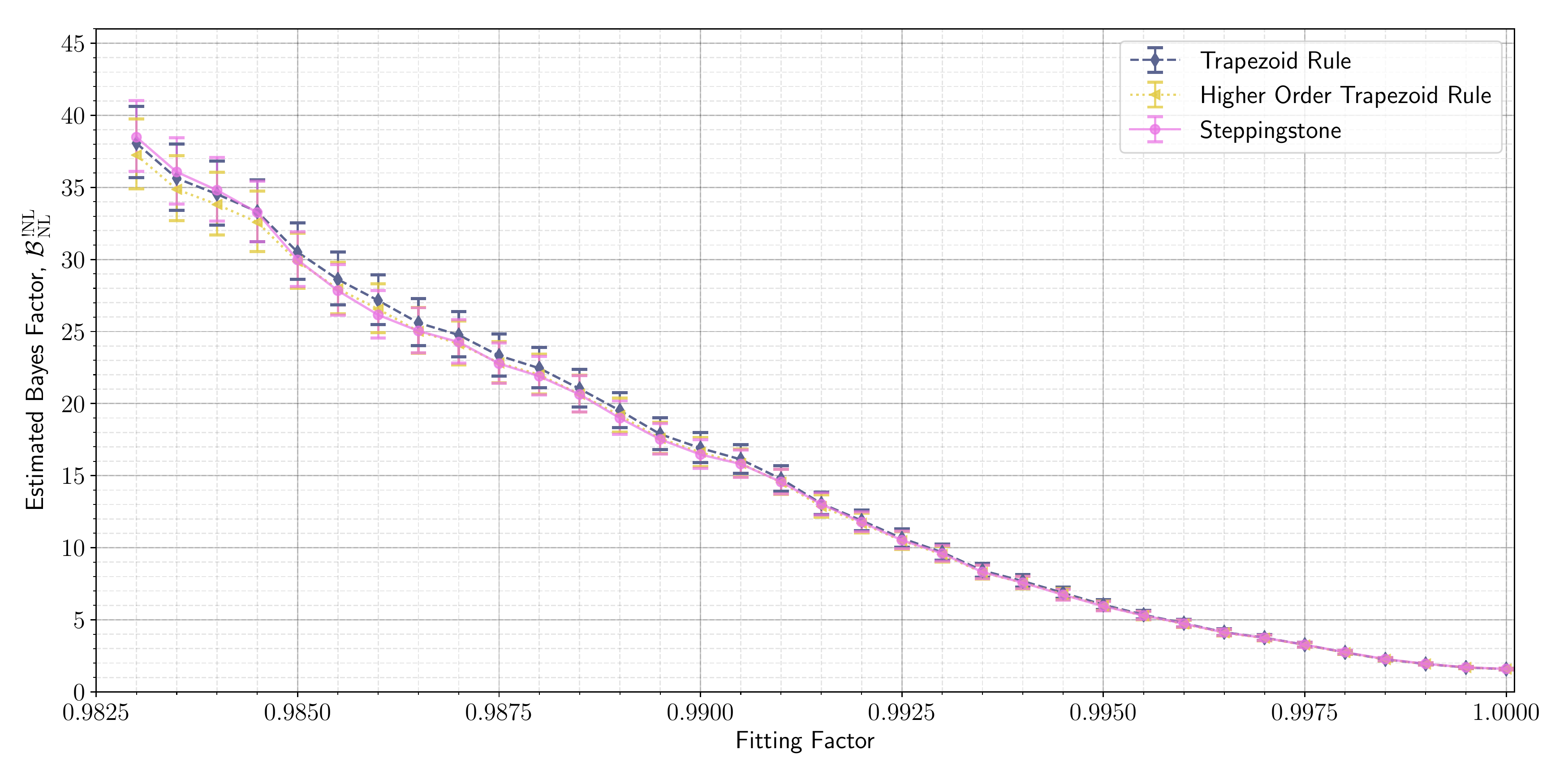}\caption{The estimated Bayes factors for nonlinear tidal parameters when the samples are filtered by the fitting factor to a non-spinning, mass-only template bank of TaylorF2 waveforms. The convention in Bayes factor is switched from the main body of the text to represent the Bayes factor for the ratio of evidence for no nonlinear tides, $p\left(\mathbf{d}\,|\, \mathrm{H}_\mathrm{TaylorF2}\right)$, to the evidence for nonlinear tides, $p\left(\mathbf{d}\,|\,\mathrm{H}_\mathrm{TaylorF2+NL}\right)$. This is abbreviated as $\mathcal{B}^{\,\mathrm{!NL}}_{\,\mathrm{NL}}$. The three methods for estimating the Bayes factor are the thermodynamic integration method from trapezoid rule integration (dark grey, dashed line), the thermodynamic integration method from the higher order trapezoid rule (yellow, small-dashed line), and the steppingstone algorithm (dark pink, solid line). A bootstrap method is used to estimate approximate errors on the Bayes Factors. Error bars represent $5th$ and $95th$ percentiles. The sampling error becomes large at a fitting factor $\lesssim 99$\%.
}
\label{fig:bayes_ff_plot}
\end{figure}

Given the observed parameter degeneracies, we now investigate regions of the parameter space where nonlinear tidal effects are not degenerate with standard waveforms by thresholding the prior distribution of $p$-$g$ waveforms on their fitting factor with standard waveforms. We combine the results of our analysis on the uniform mass, $\delta \phi$ constrained, narrow $f_0$ prior distribution model to obtain $22,600$ independent samples. We then examine the fitting factor of every independent sample, from every temperature, with a non-spinning, mass-only template bank of TaylorF2 waveforms with comparable masses to GW170817. For simplicity, we only keep the mass parameters and $p$-$g$ mode parameters in the overlap calculations, since the correlation between nonlinear tidal dynamics is most apparent in the measured chirp mass. When we examine the fitting factor between nonlinear tidal waveforms and this template bank we observe that there is a very high match between standard templates and nonlinear tidal waveforms when $n = 4/3$. The nonlinear tidal waveforms that least match this template bank tend to be those parameterized by large amplitude and large gravitational-wave phase shift. We then recompute the Bayes factor when discarding samples from the analysis below a particular fitting factor with the template bank. To ensure a robustness of the point-estimate we use a bootstrap method to estimate the Monte Carlo error for this Bayes factor estimate~\citep{efron1992bootstrap}. The bootstrap estimated Monte Carlo error tends to be much larger than the convergence error for this analysis and so we neglect inclusion of convergence error in the estimate. A statistically significant Bayes factor of $\sim 30 \, (20)$, against nonlinear tides, is found when the waveform has an overlap less than $98.5 \, (98.85)$\% with the standard waveform, see Fig.~\ref{fig:bayes_ff_plot}. While this metric is insufficient to rule out the $p$-$g$ mode instability, it is a useful metric in understanding why the evidence is nearly identical to the evidence from~\cite{de2018tidal}. We find that portions of the $p$-$g$ mode parameter space that most contribute towards the evidence come from regions of the parameter space that have a high overlap with standard waveforms. This occurs either through $A$ being too small to induce a large change in the phase of the waveform or through an associated parameter degeneracy with the chirp mass caused by large $A$, low $f_0$, and $n \sim 4/3$.

Finally, we examine the leading order estimated energy dissipated through nonlinear tides for the case of a uniform prior on the mass, with $15 \leq f_0 \leq 100$~Hz, with a $\delta \phi$ $>$ $0.1$ radian constraint. In our analysis, the $95^{\mathrm{th}}$ percentile of the estimated energy dissipated through nonlinear tides from our prior distribution is approximately $2.6 \times 10^{51}$~ergs at the terminating frequency of the TaylorF2 waveform, $f_\mathrm{ISCO}$. The estimated energy radiated by gravitational waves by neutron stars of the estimated mass range of GW170817 is greater than $\sim 10^{53}$~ergs. Our analysis finds the energy dissipated through nonlinear tides at the $95\%$ posterior credible percentile is $3 \times 10^{50}$~ergs. We find our $95\%$ posterior credible percentile to be less than the $90\%$ confidence interval constraint of $\lesssim 2.7 \times 10^{51}$~ergs in~\cite{abbott2019constraining}. Samples from our posterior distribution that have dissipation energies greater than the $90$\% credible interval tend to come from two modes in the parameter space. The first mode is from parts of the parameter space with large $A$, for $n \sim 4/3$, low $f_0$, and $\delta \phi \sim 100$~rad. The second mode is from parts of the parameter space with $A \gtrsim 10^{-8}$, for $1.6 \lesssim n < 3.0$, and $\delta \phi \sim 1-10$~rad. The high end of the nonlinear tidal energy constraints are thus dominated by waveforms that are degenerate with the standard signal.

\section{Discussion}
In this paper, we have used the observation of GW170817 and the model of~\cite{Essick:2016tkn} to look for evidence of nonlinear tides from $p$-$g$ mode coupling during the inspiral~\citep{Weinberg:2013pbi,Weinberg:2015pxa,Zhou:2018tvc}. Over the broad prior space, we find a Bayes factor of unity which gives an inconclusive result on whether nonlinear tides are favored or disfavored in GW170817, consistent with \cite{abbott2019constraining}. This Bayes factor can be interpreted as stating that there is insufficient evidence to change our prior beliefs about the credibility of the $p$-$g$ mode hypothesis after the observation of GW170817. A closer examination of the posterior distribution lead us to conclude that nonlinear tides are consistent with the signal GW170817 because they either cause very small phase shifts to the waveform, or the nonlinear tides must enter the waveform in a way that is degenerate with the other intrinsic parameters of GW170817. Regions of the nonlinear tide parameter space that have a fitting factor of less than 99\% (98.5\%) are  disfavored by a Bayes factor of 15 (25). We find that waveforms from a $p$-$g$ mode instability with overlap $>98.5$ \%, tend to either induce a very small phase shifts to the waveform or are degenerate with other intrinsic parameters of GW170817. This leads us to conclude that modeling GW170817 with nonlinear tidal parameters may not offer advantages over using a simpler model. We conclude that the consistency of the GW170817 signal with the model of \cite{Essick:2016tkn} is due to parameter degeneracy and that regions where nonlinear tides produce a measurable effect are strongly disfavored.

In principle, one could improve our analysis by separately parameterizing the amplitude, turn-on frequency, and frequency evolution for each star as in~\cite{abbott2019constraining}. However, we find our results to be broadly consistent with~\cite{abbott2019constraining}, and so we do not expect these to affect the main conclusion of our paper. Further improvements to the parametric model of $p$-$g$ mode instability could include a higher order post-Newtonian expansion of the instability, or further understanding of the instability's interaction with neutron star magnetic fields~\citep{Weinberg:2015pxa}. Nonlinear tides are poorly understood and the contribution from other stellar oscillation modes may yet contribute to a more accurate picture of the interior dynamics of neutron stars~\citep{Andersson:2017iav}. Current models of the gravitational-wave phase shift caused by nonlinear tides from the $p$-$g$ mode instability suffer from parameter degeneracies with the other intrinsic parameters of a neutron star binary.  A measurement of the binary's chirp mass that is independent of gravitational-wave observations would break this degeneracy. However, for a system like GW170817, this would require measurement of the binary's chirp mass to a precision greater than $\sim 0.02 \%$ using an electromagnetic counterpart, which is implausible. Absent improved theoretical understanding of nonlinear tides from $p$-$g$ mode coupling, it is unlikely that future observational constraints will be able to significantly improve our knowledge of these physical processes.


\acknowledgments

We thank Reed Essick, and Nevin Weinberg for helpful discussion and pointing out errors in our Bayes factor calculation in an earlier draft of this manuscript~\citep{Essick:2018wvj}. We thank Chaitanya Afle, Nils Andersson, Soumi De, Daniel Finstad, and Pantelis Pnigouras for helpful discussions. We thank Alex Nitz for writing the initial version of the code for nonlinear tides in PyCBC. The authors were supported by the National Science Foundation grant PHY-1707954. Computational work was supported by Syracuse University and National Science Foundation grant OAC-1541396. This research has made use of data obtained from the Gravitational Wave Open Science Center (\url{https://www.gw-openscience.org/about/}).

\software{PyCBC Inference \citep{alex_nitz_2018_1208115,biwer2019pycbc},  
          emcee \citep{emcee,vousden:2016}, 
          LIGO Algorithm Library \citep{lal},
          Matplotlib \citep{Hunter:2007},
          Scipy \citep{scipy}
          }


\end{document}